\documentstyle[preprint,eqsecnum,aps]{revtex}

\begin{document}
\draft
\title{Haldane Exclusion Statistics and the Boltzmann Equation}
\author{R.K. Bhaduri$^1$, R.S. Bhalerao$^2$ and M.V.N. Murthy$^3$}
\address{
$^1$ Department of Physics and Astronomy, McMaster University,\\
Hamilton, Ontario, L8S 4M1, Canada\\
\bigskip
$^2$ Theoretical Physics Group,\\
Tata Institute of Fundamental Research,\\
Homi Bhabha Road, Colaba, Bombay 400 005, India\\
\bigskip
$^3$ Institute of Mathematical Sciences, Taramani,\\
Madras 600 113, India
}
\date{\today}
\maketitle
\begin{abstract}
We generalize the collision term in the one-dimensional
Boltzmann-Nordheim transport
equation  for quasiparticles that obey the Haldane exclusion
statistics. For the equilibrium situation, this
leads to the ``golden rule'' factor for quantum transitions. As an
application of this, we calculate the density response function of a
one-dimensional electron gas in a periodic potential, assuming that
the particle-hole excitations are quasiparticles obeying the new
statistics. We also calculate the relaxation time of a nuclear spin in a metal
 using the modified golden rule.
\end{abstract}

\newpage
\begin{center}
\bf {1.~~~INTRODUCTION}
\end{center}
\bigskip

Recently, Haldane has proposed a new exclusion statistics for
confined quasiparticles, that interpolates between the Fermi and Bose
quantum behaviours, and gives rise to partial blocking of a singly
occupied state.$^{(1)}$ There has been considerable theoretical
activity concerning study of statistical and thermodynamical
properties of systems of particles obeying the generalised exclusion
statistics. Examples of quasiparticles obeying this new statistics
include spinons in antiferromagnetic spin chain with inverse square
exchange interaction$^{(2)}$, and two-dimensional anyons restricted
to the lowest Landau level in a strong magnetic field.$^{(3)-(5)}$
Other examples are the Calogero-Sutherland model$^{(6)-(7)}$ in one
dimension, and particles interacting with a two-body delta-function
potential in two-dimensions in the thermodynamic limit.$^{(8)}$
Haldane$^{(9)}$ has analysed a Heisenberg spin chain in one-dimension
with inverse-square coupling, and shown that under certain conditions
the quasi-particle excitations obey a particularly simple form of the
generalised statistics. Despite all these theoretical studies, it
is not entirely clear under what conditions a physical or chemical
system will experimentally manifest this generalised statistics. The
best evidence so far comes from recent neutron inelastic scattering
experiment$^{(10)}$ on the compound $KCuF_3$ which is a
one-dimensional Heisenberg antiferromagnet above $40^0 K$. The
observed inelastic scattering is best fitted by spinon excitations in
a spin chain whose interactions fall off as the inverse square of the
lattice distance$^{(11)}$. The dynamic spin correlation function for such
 a system has been calculated by Haldane and Zirnbauer.$^{(12)}$

The present paper addresses a question of current interest,
namely how to calculate transport properties of particles obeying the
generalised exclusion statistics. Specifically, how does the Boltzmann
transport equation get modified in such a situation? It may be
recalled that the collision term in the Boltzmann equation was
modified by Nordheim$^{(13)}$ to incorporate the effect of quantum
statistics. Account was taken of the Pauli blocking of the final state
in binary collision events for particles obeying Fermi statistics. On
the other hand, an enhancement factor was introduced (as in the
Einstein coefficient of induced emission$^{(14)}$) in the final state
occupancy for Bose statistics. The suppression and enhancement
factors, $(1-f)$ and $(1+f)$, appropriate for Fermions and Bosons,
respectively, are routinely used in many-body calculations. What are
the corresponding factors for the systems of particles obeying the new
statistics? We have presented the answer to this question. Future
calculations of transport properties of such systems will find these
results relevant. Besides generalizing the Boltzmann-Nordheim
equation, we have also generalized the Fermi Golden Rule.

The plan of this paper is as follows. In section 2, we examine changes
in the structure of the collision term in the Boltzmann equation due
to the incorporation of the new statistics. This yields the ``golden
rule'' factor for quantum transitions. As an application of this, we
examine, in section 3, the response of a one-dimensional electron gas
to an external periodic potential, and calculate its density response
function. Whereas the ground state of the electron gas is taken to be
the conventional Fermi system, we assume that the particle-hole
excitations are quasiparticles obeying the new statistics. The
magnitude of the density response function is then shown to be
appreciably enhanced, thereby lowering the Peierls transition
temperature. In section 4, we generalize the Fermi Golden Rule for
quasiparticles obeying the new statistics and apply it to the
calculation of a relaxation time of a nuclear spin in a metal. Both
the applications are schematic, with a view to explore qualitative
changes from the standard results.

\newpage
\begin{center}
\bf {2.~~~GENERALIZED BOLTZMANN-NORDHEIM EQUATION}
\end{center}
\bigskip

We now derive the Boltzmann equation (specifically, the collision
term) for particles obeying Haldane's statistics.  For the derivation
of the collision term, we make the usual assumptions of (i) the
dominance of binary collisions, (ii) the hypothesis of molecular chaos
(``stosszahlansatz''), and (iii) the slow variation of the
distribution function $f(x,p,t)$ over distances and times of the order
of characteristic interaction lengths and durations, respectively.
Recall, $f(x,p,t) dx~dp / (2 \pi)$ is the number of particles in the
phase-space volume element $dx~dp$ at time $t$. (We set $\hbar = 1$
throughout this paper.) We wish to derive an expression for the
collision term $C$ in the Boltzmann equation:
\begin{equation}
{\partial f \over \partial t} + v {\partial f \over \partial
x} + {\cal F} {\partial f \over \partial p} = C.
\eqnum{1}
\end{equation}
$C$ is the rate of change of $f$ caused by collisions.  Collisions of
the type $p' p'_1 \rightarrow p p_1$ tend to increase the population
of particles having momentum $p$, while those of the type $pp_1
\rightarrow p' p'_1$ tend to decrease it.  Let $w(p'p'_1 \rightarrow
pp_1) dp ~dp_1$ be the transition probability per unit volume and per
unit time that two particles having incoming momenta $p'$ and $p'_1$
are scattered with outgoing momenta in the ranges $(p,p+dp)$ and
$(p_1,p_1+dp_1)$.  Then according to the hypothesis of molecular
chaos, for {\it classical} (Maxwell-Boltzmann (MB)) particles the net
increase in the number of particles in $dx~dp$ due to collisions
occurring during $dt$ is$^{(15)}$
\begin{eqnarray}
\int \Big[f' dp'.f'_1dp'_1 . w(p'p'_1 &\rightarrow& pp_1)dp~dp_1~dx~dt
\nonumber\\
-f~dp . f_1 dp_1 . w(pp_1 &\rightarrow& p'p'_1) dp' ~dp'_1 ~dx~dt\Big],
\nonumber
\end{eqnarray}
where $f' \equiv f(x,p',t)$, etc., and the integration is only over
$dp'dp'_1dp_1$.  Using the detailed-balance property $w(p'p'_1
\rightarrow pp_1) = w(pp_1 \rightarrow p'p'_1)$ and inserting a factor
${1\over2}$ to take into account the fact that a final state with
momenta $(p',p'_1)$ is indistinguishable from that with momenta
$(p'_1,p')$ we get
\begin{equation}
C = {1\over 2} \int dp_1 dp' dp'_1 ~w(pp_1 \rightarrow p'p'_1) (f' f'_1 -
ff_1).
\eqnum{2}
\end{equation}

For particles obeying {\it quantum} statistics, however, the above
derivation needs to be modified because now the transition probability
depends also on the occupancy of the final state.  Thus, $(f'f'_1 -
ff_1)$ in (2) should be replaced by $f'f'_1 ~F(f) F(f_1) - ff_1
{}~F(f')F(f'_1)$, where $F$ takes account of the effect of the
quantum statistics on the accessibility of the final state. For
example, in the case of Fermi-Dirac (FD) statistics, $F(f) = (1 - f)$
to incorporate Pauli blocking, and for Bose-Einstein (BE) statistics,
$F(f) = (1 + f)$ to take care of Einstein enhancement. According to
the $H$-theorem in statistical mechanics, the entropy production
vanishes if and only if
\begin{equation}
f'f'_1 ~F(f) F(f_1) = ff_1 ~F(f') F(f'_1).
\eqnum{3}
\end{equation}
Solution of this equation provides the shape of the equilibrium
distribution function $f_{eq}$.  Conversely, as we proceed to show, if
$f_{eq}$ is known, (3) can be used to deduce the functional form
$F(f)$. Rewriting (3) in the form
\[
{f \over F(f)} . {f_1 \over F(f_1)} = {f' \over F(f')} . {f'_1 \over
F(f'_1)},
\]
taking logarithms of both sides, and using the fact that energy $(E)$ is
the appropriate summational invariant in the collision, we get
\[
{\rm ln}[f/F(f)] = -a(E - b),
\]
where $a$ and $b$ are constants to be identified with inverse
temperature ($T$) and chemical potential ($\mu$), respectively. Thus,
we get finally
\[
F(f) = \exp[a(E - b)] f = \zeta(E) f,~~~~~~ \zeta(E)
 \equiv \exp[(E-\mu)/T], \]
where we have set the Boltzmann constant to unity. For the $MB$, $FD$
and $BE$ statistics, the equilibrium distribution functions $f$ are
$\zeta^{-1}, 1/(\zeta+1)$ and $1/(\zeta-1)$, respectively. Hence, the
corresponding $F$'s are $1$, $(1-f)$ and $(1+f)$, as expected.

Now, for particles obeying the Haldane statistics, the equilibrium
distribution function, as shown by Wu$^{(5)}$, is
\begin{equation}
f(E) = 1/(\omega(E) + \alpha),
\eqnum{4}
\end{equation}
where $\alpha ~(0 \leq \alpha \leq 1)$ is the statistical
interpolation parameter ($\alpha = 0$ corresponds to Bosons and
$\alpha=1$ to Fermions) and $\omega(E)$ satisfies
\begin{equation}
\omega (E)^\alpha [1 + \omega (E)]^{1-\alpha} = \exp[(E-\mu)/T].
\eqnum{5}
\end{equation}
It is straightforward to show that, in this case,
\begin{equation}
F(f) = (1 - \alpha f)^\alpha [1 + (1 - \alpha) f]^{1 - \alpha}.
\eqnum{6}
\end{equation}
In particular, for semions $\left(\alpha = {1\over2}\right)$ we get
\begin{eqnarray}
f &=& 1/\left(\displaystyle {1\over4} + \zeta^2\right)^{1/2},\nonumber\\
F(f) &=& (1 - f/2)^{1/2} (1 + f/2)^{1/2}.
\eqnum{7}
\end{eqnarray}

Thus, the Boltzmann equation (1) for particles obeying Haldane's statistics
is given by
\begin{eqnarray}
{\partial f \over \partial t} &+& v {\partial f \over \partial x}
+ {\cal F} {\partial f \over \partial p}\nonumber\\
&=& {1\over2} \int dp_1 dp' dp'_1 ~w(pp_1 \rightarrow p'p'_1)
\big[f'f'_1 ~F(f) F(f_1) - ff_1 ~F(f') F(f'_1)\big],\nonumber
\end{eqnarray}
where $F$ is given by (6).

\newpage
\begin{center}
\bf {3.~~~DENSITY RESPONSE FUNCTION}
\end{center}
\bigskip

In calculating the density response function $\chi(Q)$ of the
one-dimensional electron gas, we follow the treatment given by
Kagoshima {\it et al.}$^{(16)}$ For a periodic potential $V(r) = \displaystyle
\sum_Q V_Q e^{iQr}$, the charge distribution of the electrons
in a length $L$ is distorted by $\delta
\rho(r) = {1 \over L} \displaystyle \sum_Q \rho_Q e^{iQr}$, and
$\chi(Q)$ is just the linear response function, obtained from
\[
\rho_Q = -V_Q~ \chi (Q),
\]
by calculating $\delta \rho(r)$ perturbatively. Consider an electron
in the initial state $k$, absorbing a phonon of wave number $Q ~(-Q)$,
reaching the final state $k + Q ~(k - Q)$.  Then, using $FD$
statistics, the usual expression for $\rho_Q$ is$^{(16)}$
\begin{equation}
\rho_Q = {V_Q \over N} \sum_k \left[{f_k (1 - f_{k+Q}) \over E_k - E_{k+Q}}
+ {f_k (1 - f_{k-Q}) \over E_k - E_{k-Q}}\right],
\eqnum{8}
\end{equation}
where we have neglected the phonon energy $\hbar \omega_Q$ in
comparison to the electron energy
$E_k = \hbar^2 k^2/2m$, and $N$ is the number of
electrons in length $L$. The dummy variable $k$ in the second term
may be replaced by $(k + Q)$, resulting in the cancellation of the
bilinear terms in $f$'s, and yielding
\begin{equation}
\chi (Q) = {1 \over N} \sum_k {f_{k+Q} - f_k \over E_k - E_{k+Q}}.
\eqnum{9}
\end{equation}
At temperature $T=0$, we have $f_k = 1$ and $f_{k+Q} = 0$.
In one dimension, this results in a logarithmic
singularity for transitions from $k = \pm k_F$ to $k = \mp k_F$ with
$Q = \mp 2k_F$. This is a manifestation of the well-known Peierls
instability. At nonzero temperature $T \equiv \beta^{-1}$, the
singularity is replaced by a finite result, as may be verified by a
straight-forward calculation:
\begin{equation}
\chi(Q = 2k_F) = {D(E_F) \over 2N} \int^{\epsilon_B \beta/2}_0 {\tanh x
\over x} dx,
\eqnum{10}
\end{equation}
where $D(E_F)$ is the density of states at the Fermi level and the
energy integration is carried out only in the range $|E - E_F| \leq
\epsilon_B \ll E_F$. We repeat this calculation for $\chi$ by
replacing the factors $f(1-f)$ in (8) by the new factors $fF(f)$
derived in section 2. There is of course, no a
priori justification for assuming that these excitations obey the new
statistics. Nevertheless, if they did, what would be the experimental
signals in a one-dimensional system? The enhancement of $\chi (Q)$
near $Q = 2k_F$ has a direct bearing on the estimation of the Peierls
metal-insulator transition temperature, as well as on the Kohn
anomaly. It may therefore be a worthwhile exercise to reexamine $\chi
(Q)$ with the modified golden rule.

It is simplest to illustrate the calculation with semions $(\alpha =
1/2)$, where the equilibrium distribution is explicitly known (see
(7)). The expression in the square brackets in (8) is then replaced by
\begin{equation}
\left[{f_k\left(1 - {1 \over 4}f^2_{k+Q}\right)^{1/2} \over E_k -
E_{k+Q}} + {f_k\left(1 - {1 \over 4}f^2_{k-Q}\right)^{1/2} \over E_k -
E_{k-Q}} \right].
\eqnum{11}
\end{equation}
Note that on changing the dummy variable $k \rightarrow k + Q$ in
the second term in (11), there is no cancellation of the bilinear
terms in $f$ as earlier. However, at $T=0$, the quasiparticle
excitations near the Fermi surface have $f_k = 2$, $f_{k+Q} = 0$, and
hence, there is an overall enhancement of $\chi (Q)$ by a factor of
$2$, again with a logarithmic singularity at $Q = 2k_F$. At nonzero
temperature, the semionic distribution function $f$ given by (7)
may be used to evaluate $\chi (Q)$, and one obtains after
straight-forward algebra
\begin{equation}
\chi (Q = 2 k_F) = {D(E_F) \over 2N} 2 \int^{\epsilon_B \beta /2}_0
{dx \over x} {2(e^{2x} -
e^{-2x}) \over (1 + 4e^{-4x})^{1/2} (1 + 4e^{4x})^{1/2}},
\eqnum{12}
\end{equation}
which should be compared with (10). The integrals in (10) and
(12) are plotted as a function of $\epsilon_B \beta /2$ in Fig. 1, and
again show an enhancement of $\chi (Q)$, but by a factor somewhat
greater than $2$. Although there is an enhancement in the response
function in our simple model, because the temperature dependence of it
does not change, it may not be possible to detect this effect
experimentally.

\newpage
\begin{center}
\bf {4.~~~GENERALIZED FERMI GOLDEN RULE AND RELAXATION TIME}
\end{center}
\bigskip

We now consider generalization of the well-known Fermi Golden Rule,
for particles obeying the fractional exclusion statistics.
Consider a two-particle scattering process $k + p
\rightarrow k'+p'$, in one dimension. Let the two particles be
distinct. We have in mind processes like the electron-nuclear
interactions in metals. According to the Fermi Golden Rule, the
transition probability per unit time for such a process is given by
\[
w_{i\rightarrow f} = 2 \pi |<i| H_{int}|f>|^2 \delta(E_i-E_f),
\]
where $H_{int}$ is the interaction Hamiltonian and the delta function
ensures energy conservation between the initial and the final states.
Typically in electron-nuclear interactions (as in magnetic relaxation
in solids) one can neglect the nuclear recoil to a very good
approximation and therefore one can replace $\delta(E_i-E_f)$ by
$\delta (E_k -E_{k'})$, $k$ and $k'$ being the initial and final
momenta of the scattered electron. We work in this limit to keep
the calculation
simple and also because this is the relevant limit for many physical
applications. The total transition probability per unit time at
nonzero temperature is then given by$^{(17)}$
\[
W_{i\rightarrow f} = 2 \pi \int dk dk'
|<i|H_{int}|f>|^2  \delta(E_k-E_{k'}) \rho(k) \rho(k') f(E_k)
F(E_{k'}),
\]
where $\rho(k)$ denotes the density of states in the momentum space,
$f(E)$ is as in (4) and $F(E)$ is what was denoted by $F(f)$
earlier, see (6). Performing one of the two integrations by using
the delta function, we obtain
\begin{equation}
W_{i\rightarrow f} = 2 \pi \int dE
|<i|H_{int}|f>|^2 (\rho(E))^2 f(E) F(E).
\eqnum{13}
\end{equation}

In order to evaluate the integral in (13), we note that
\begin{equation}
f(E)F(E) = (f(E))^2~ e^{(E-E_F)/T},
\eqnum{14}
\end{equation}
where $E_F$ is the Fermi energy. Notice that in the low-temperature
limit we have
\[
f(E) = 1/\alpha ~~~~~~ E \le E_F,
\]
and zero otherwise. Thus at low enough temperatures, the system for an
arbitrary $\alpha$ (except very close to the Bosonic end) does exhibit
a Fermi surface. We will be using this fact later in the
calculation. Now for reasons that will become clear shortly, we wish
to obtain an expression for the derivative $df/dE$. Using (4-5),
we get
\begin{equation}
\frac {df}{dE} = - (f(E))^2 \frac {d\omega}{dE}
= - (f(E))^2 \frac {\omega(1 + \omega)}{T (\alpha + \omega)}.
\eqnum{15}
\end{equation}
Substituting $(f(E))^2$ from (15) into (14), we get
\begin{equation}
f(E) F(E) = - \frac{df}{dE} \frac{T(\alpha + \omega)}{\omega (1 +
\omega)} e^{(E - E_F)/T}.
\eqnum{16}
\end{equation}

We now consider the low-temperature ($T \ll E_F$) limit of the process
under consideration. In this limit, $ f(E) \approx \theta(E_F-E) /
\alpha,$ and hence, $df/dE \approx - \delta(E_F-E) / \alpha$.
Substituting this in (16), we find
\begin{equation}
f(E)F(E) \approx \frac {T(\alpha+\omega(E))}{\alpha
\omega(E)(1+\omega(E))} \delta(E_F-E).
\eqnum{17}
\end{equation}

Substituting (17) in (13) and performing the energy integration
we get the Generalized Fermi Golden Rule:
\[
W_{i \rightarrow f} = 2 \pi
|<i|H_{int}|f>|^2 (\rho (E_F))^2 \frac {T(\alpha + \omega (E_F))}
{\alpha \omega (E_F)(1 + \omega (E_F))}.
\]
In the special case when $\alpha = 1$ (Fermions), we have, from (5),
$\omega(E_F) = 1$, and we get
\[
W^F_{i\rightarrow f} = 2 \pi
|<i|H_{int}|f>|^2 (\rho(E_F))^2 T.
\]
We can therefore, in general, write
\[
W^{\alpha}_{i \rightarrow f} = W^F_{i \rightarrow f}
\frac {\alpha + \omega(E_F)} {\alpha \omega(E_F)(1 + \omega(E_F))}.
\]

We may now apply this result to specific cases. A straightforward
application is to the calculation of a relaxation time of a
nuclear spin in a metal. The relaxation time for
arbitrary $\alpha$ is then given by$^{(17)}$
\[
\frac{1}{\tau}(\alpha)  = \frac{1}{\tau}(Fermions)
\frac {\alpha + \omega(E_F)}{\alpha \omega(E_F)(1+\omega(E_F))}.
\]
Thus the change due to fractional statistics is simply given by a
multiplicative factor which depends on $\alpha$. In particular, at
$\alpha =1/2$, this multiplicative factor can be explicitly calculated,
and we get
\[
\frac{1}{\tau}(\alpha)  =\sqrt{5} \frac{1}{\tau}(Fermions).
\]

Notice that in deriving the Generalized Fermi Golden Rule we have made
a number of simplifying assumptions. Real systems are likely to be
more complicated. Nevertheless, the above derivation probably
indicates the correct direction in which the transition rates move
when fractional exclusion statistics particles are involved.

In summary, we have generalized the Boltzmann-Nordheim equation and
the Fermi Golden Rule for quasiparticles obeying the Haldane exclusion
statistics. As two simplified
applications of these results, we have calculated the density response
function and the relaxation time of a one-dimensional gas obeying the
new statistics. Although the applications presented here are simple
and exploratory, we believe that they are the first attempts in this
area to make a connection of theory to the consideration of observable
effects. Moreover, this could be attempted only because the transport
equation was nontrivially modified.

\acknowledgments

We thank G. Rajasekaran, Diptiman Sen and R. Shankar for discussions.
One of us (RKB) would like to acknowledge the hospitality of the Tata
Institute of Fundamental Research where a part of this work was
done. RKB also thanks the TOKTEN program of UNDP for financial
support.

\newpage
\begin{center}
\bf {REFERENCES}
\end{center}
\bigskip

\begin{enumerate}

\item[{1.}] F.D.M. Haldane, {\it Phys. Rev. Lett.} {\bf 67}: 937 (1991).

\item[{2.}] F.D.M. Haldane, {\it Phys. Rev. Lett.} {\bf 60}: 635 (1988);
B.S. Shastry, {\it ibid.} {\bf 60}: 639 (1988).

\item[{3.}] M.V.N. Murthy and R. Shankar, {\it Phys. Rev. Lett.} {\bf 72}:
3629 (1994).

\item[{4.}] A. Dasni\`eres de Veigy and S. Ouvry, {\it Phys. Rev. Lett.}
{\bf 72}: 600 (1994).

\item[{5.}] Y.-S. Wu, {\it Phys. Rev. Lett.} {\bf 73}: 922 (1994).

\item[{6.}] F. Calogero, {\it J. Math. Phys.} {\bf 10}: 2197 (1969).

\item[{7.}] B. Sutherland, {\it J. Math. Phys.} {\bf 12}: 246 (1971);
{\it ibid.} {\bf 12}: 251 (1971); {\it Phys. Rev. A} {\bf 4}: 2019 (1971).

\item[{8.}] R.K. Bhaduri and M.K. Srivastava, McMaster University preprint,
July (1995).

\item[{9.}] F.D.M. Haldane, {\it Phys. Rev. Lett.} {\bf 66}: 1529 (1991).

\item[{10.}] R.A. Cowley, D.A. Tennant, S.E. Nagler
and T. Perring, {\it J. Magnetism and Magnetic Materials} {\bf 140-144}:
1651 (1995).

\item[{11.}] G. Mueller, H. Thomas, H. Beck and J.C. Bonner,
{\it Phys. Rev. B} {\bf 24}: 1429 (1981).

\item[{12.}] F.D.M. Haldane and M.R. Zirnbauer,
{\it Phys. Rev. Lett.} {\bf 71}: 4055 (1993).

\item[{13.}] L.W. Nordheim, {\it Proc. Roy. Soc. London Ser. A} {\bf 199}:
689 (1928).

\item[{14.}] See, for example, J. J. Sakurai,
{\it Advanced Quantum Mechanics} (Benjamin, Menlo Park, 1984), page 38.

\item[{15.}] S.R. de Groot, W.A. van Leeuwen and Ch.G. van Weert,
{\it Relativistic Kinetic Theory} (North-Holland, Amsterdam, 1980).

\item[{16.}] S. Kagoshima, H. Nagasawa and T. Sambongi, {\it One-Dimensional
Conductors} (Springer-Verlag, Berlin, 1988), chap. 2.

\item[{17.}] A. Abragam, {\it The Principles of Nuclear Magnetism}
(Oxford, London, 1962), page 358.

\end{enumerate}

\newpage
\begin{center}
\bf{FIGURE CAPTION}
\end{center}
\begin{enumerate}

\item[{Fig. 1}]: Dashed and solid lines represent integrals occurring
in (10) and (12) respectively, as a function of $\epsilon_B
\beta/2$, showing an enhancement of the density response function
$\chi(Q = 2 k_F)$ at various temperatures, as one goes from Fermionic
to semionic statistics.

\end{enumerate}
\end{document}